\documentclass[aps,preprint, groupedaddress, floatfix, nofootinbib]{revtex4-1}

\usepackage{changes}
\usepackage{epsfig}
\usepackage{graphicx}
\usepackage{amsmath}
\usepackage{bbold}
\usepackage{tabularx, booktabs}
\usepackage[normalem]{ulem}
\usepackage{soul}  
\usepackage{xcolor} 
\usepackage{multirow} 
\usepackage{xr}

\begin{document}

\title{A large modulation of electron-phonon coupling and an emergent
  superconducting dome in doped strong ferroelectrics}

\author{Jiaji Ma$^1$}
\author{Ruihan Yang$^1$}
\author{Hanghui Chen$^{1,2}$}
\affiliation{
$^1$NYU-ECNU Institute of Physics, NYU Shanghai, Shanghai 200122, China\\
$^2$Department of Physics, New York University, New York, NY 10003, USA\\
{ $^*$email: hanghui.chen@nyu.edu}}

\date{\today}
\begin{abstract}
We use first-principles methods to study doped strong ferroelectrics
(taking BaTiO$_3$ as a prototype). {   Here we find a strong
  coupling between itinerant electrons and soft polar phonons in doped
  BaTiO$_3$, contrary to Anderson/Blount's
weakly coupled electron mechanism for ``ferroelectric-like metals''.}
As a consequence, across a
polar-to-centrosymmetric phase transition in doped BaTiO$_3$, the
total electron-phonon coupling is increased to about 0.6 around the
critical concentration, which is sufficient to induce phonon-mediated
superconductivity of about 2 K. Lowering the crystal symmetry of doped
BaTiO$_3$ by imposing epitaxial strain can further increase the
superconducting temperature via a sizable coupling between itinerant
electrons and acoustic phonons. Our work demonstrates a
viable approach to modulating electron-phonon coupling
and inducing phonon-mediated superconductivity in doped strong
ferroelectrics and potentially in polar metals. Our results also show
that the weakly coupled electron mechanism for ``ferroelectric-like
metals'' is not necessarily present in doped strong ferroelectrics.
\end{abstract}
\maketitle

\section*{Introduction}
Electron-phonon coupling plays an important role in a variety of
physical phenomena in solids~\cite{Grimvall}. In metals and doped
semiconductors, low-energy electronic excitations are strongly
modified by the coupling of itinerant electrons to lattice vibrations,
which influences their transport and thermodynamic
properties~\cite{Ziman}. Furthermore, electron-phonon coupling
provides an attractive electron-electron interaction, which leads to
conventional (i.e. phonon-mediated) superconductivity in many
metals~\cite{BCS}. Recent studies on hydrogen-rich materials show that
when their electron-phonon coupling is strong enough, the transition
temperature of conventional superconductors can reach as high as 260 K
at 180-200 GPa~\cite{SC260, SC203, SC250}. One general way to increase
the electron-phonon coupling of solids is to find a particular phonon
to which itinerant electrons are strongly coupled and whose softening
(i.e. the phonon frequency approaches zero) across a structural phase
transition may consequently increase the total electron-phonon
coupling~\cite{ALLEN19831}.  However, identifying a strong coupling
between a soft phonon and itinerant electrons in real materials is no
easy task, which relies on material details. On the other hand, the
superconductivity in doped SrTiO$_3$ has drawn great interests from
both
theorists~\cite{Kozii2019,PhysRevB.97.144506,PhysRevLett.115.247002,PhysRevB.98.104505,PhysRevB.94.224515,Gorkov4646,PhysRevB.100.094504,PhysRevResearch.1.013003}
and
experimentalists~\cite{PhysRevLett.112.207002,PhysRev.163.380,Russell2019,BARATOFF19811335,Rischau2017,Swartz1475,Ahadieaaw0120,Russell2019,PhysRevMaterials.3.124801,Stucky2016}. One
beautiful experiment is Sr$_{1-x}$Ca$_x$TiO$_{3-\delta}$ in which Ca
doping leads to a weak ferroelectric distortion in SrTiO$_3$ and
oxygen vacancies provide itinerant
electrons~\cite{Rischau2017,Wang2019,PhysRevLett.52.2289}. Increasing the carrier concentration in
Sr$_{1-x}$Ca$_x$TiO$_{3-\delta}$ induces a polar-to-centrosymmetric
phase transition and a superconducting ``dome'' emerges around the
critical concentration. The nature of the superconductivity in doped
SrTiO$_3$ is highly
debatable~\cite{Kozii2019,PhysRevB.97.144506,PhysRevLett.115.247002,Gorkov4646,PhysRevB.98.104505,PhysRevB.93.184507,PhysRevB.100.094504,PhysRevB.94.224515,PhysRevResearch.1.013003,Swartz1475,PhysRevLett.112.207002,PhysRev.163.380,Russell2019,BARATOFF19811335,Rischau2017},
because the superconductivity in doped SrTiO$_3$ can persist to very
low carrier
density~\cite{PhysRevX.3.021002,PhysRevB.98.104505,PhysRevB.94.224515},
which seriously challenges the standard phonon pairing
mechanism~\cite{Migdal}. {   It is not clear why superconductivity in
  doped SrTiO$_3$ vanishes above a critical concentration in spite of an
increasing density of states at the Fermi level~\cite{Tomioka2019}}.
Attention has been paid to recent proposals
on soft polar phonons, but the coupling details and strength are
controversial~\cite{PhysRevLett.115.247002,PhysRevB.98.104505,PhysRevB.93.184507,PhysRevB.100.226501,PhysRevResearch.1.013003}.
Furthermore, according to Anderson and Blount's
original proposal that inversion symmetry breaking by collective polar
displacements in metals relies on the weak coupling between itinerant
electrons and soft phonons responsible for inversion symmetry
breaking~\cite{PhysRevLett.14.217,Laurita2019,Puggioni2014}, it is not obvious that across the
polar-to-centrosymmetric phase transition the soft polar phonons can
be coupled to itinerant electrons in Sr$_{1-x}$Ca$_x$TiO$_{3-\delta}$,
or more generally in doped ferroelectrics and polar
metals~\cite{PhysRevResearch.1.013003,PhysRevB.98.104505,PhysRevB.100.226501,PhysRevB.100.226502}.

Motivated by the above experiments and theories, we use
first-principle methods with no adjustable parameters to demonstrate a
large modulation of electron-phonon coupling in doped strong
ferroelectrics by utilizing soft polar phonons. We study BaTiO$_3$ as
a prototype, because 1) previous studies found that in $n$-doped
BaTiO$_3$, increasing the carrier density gradually reduces its polar
distortions and induces a continuous polar-to-centrosymmetric phase
transition~\cite{PhysRevLett.109.247601,
  Chengliang2019}; and 2) the critical concentration for the phase
transition is about $10^{21}$/cm$^3$, which is high enough so that the
electron-phonon coupling can be directly calculated within the
Migdal's approximation (in contrast, in doped SrTiO$_3$,
superconductivity emerges at a much lower carrier concentration
$10^{17}$-$10^{20}$ /cm$^3$ {  so that its Debye frequency is
  comparable to or even higher than the
  Fermi energy $\hbar \omega_{D}/\epsilon_F \sim 1-10^2$~\cite{GASTIASORO2020168107},}
which invalidates the Migdal's
approximation and Eliashberg equation)~\cite{Migdal}.  The key result
from our calculation is that, contrary to Anderson/Blount's argument
for ``ferroelectric-like
metals''~\cite{PhysRevLett.14.217,Laurita2019,Puggioni2014}, we find
that the phonon bands associated with the soft polar optical phonons
are strongly coupled to itinerant electrons across the
polar-to-centrosymmetric phase transition in doped BaTiO$_3$. As a
consequence, the total electron-phonon coupling of doped BaTiO$_3$ can
be substantially modulated via carrier density and in particular is
increased to about 0.6 around the critical concentration. Eliashberg
equation calculations find that such an electron-phonon coupling is
sufficiently large to induce phonon-mediated superconductivity of
about 2K.  In addition, we find that close to the critical
concentration, lowering the crystal symmetry of doped BaTiO$_3$ by
imposing epitaxial strain further increases the superconducting
temperature via a sizable coupling between itinerant electrons and
acoustic phonon bands.

While ferroelectricity and superconductivity have little in common,
our work demonstrates an experimentally viable approach to modulating
electron-phonon coupling and inducing phonon-mediated
superconductivity in doped strong ferroelectrics and potentially in
polar metals~\cite{PhysRevLett.14.217, Shi2013}. Our results show that
the weakly coupled electron mechanism in ``ferroelectric-like metals''
is not necessarily present in doped strong ferroelectrics and as a
consequence, the soft polar phonons can be utilized to induce
phonon-mediated superconductivity across a structural phase
transition.

\section*{Results}

\subsection*{Structural phase transition induced by electron doping}

In this study, electron
doping in BaTiO$_3$ is achieved by adding extra electrons to the
system with the same amount of uniform positive charges in the
background. For benchmarking, our calculation of the undoped
tetragonal BaTiO$_3$ gives the lattice constant $a = 3.930$ \AA~and
$c/a=1.012$, polarization $P= 0.26$ $\rm{C/m^2}$, and Ti-O and Ba-O
relative displacements of 0.105 \AA~and 0.083 \AA, respectively,
consistent with the previous calculations
~\cite{structure,PhysRevB.56.1625,PhysRevB.96.035143}.  We note that
upon electron doping, BaTiO$_3$ becomes metallic and its polarization is
ill-defined~\cite{Resta1994}. Therefore, we focus on analyzing ionic
polar displacements and $c/a$ ratio to identify the critical
concentration~\cite{PhysRevLett.109.247601}.

\begin{figure}[t!]
\includegraphics[width=0.9\textwidth]{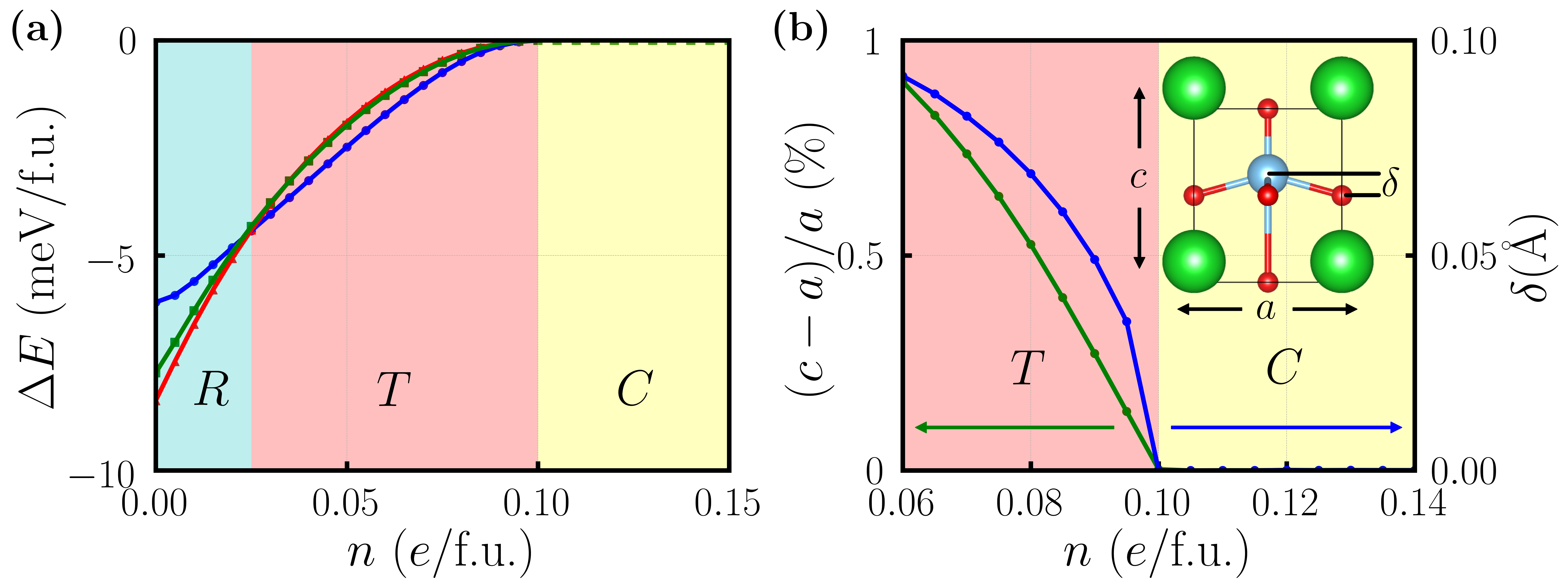}
\caption{\label{fig:structure}\textbf{ Structural phase transition induced by electron doping.}  \textbf{a}) Total energies of $n$-doped
  BaTiO$_3$ in different crystal structures: the rhombohedral
  structure ($R$, red line), the orthorhombic structure ($O$, green
  line), the tetragonal structure ($T$, blue line) and the cubic
  structure ($C$, setting as the zero point at each electron doping
  concentration $n$).
  Upon electron doping, the ground state structure of BaTiO$_3$ changes
  from $R$ to $T$, finally to $C$. \textbf{b}) The $c/a$ ratio and
  Ti-O cation displacement
  $\delta$ of $n$-doped BaTiO$_3$. $T$ means the tetragonal structure
  and $C$ means the cubic structure.  The inset shows the tetragonal structure
  of doped BaTiO$_3$ where $c$ is the long cell axis and
  $a$ is the short cell axis. $\delta$ is the displacement of the Ti
  atom with respect to the O atom layer along the $c$ axis.}
\end{figure}

We test four different crystal structures of BaTiO$_3$ with electron
doping: the rhombohedral structure (space group $R3m$ with Ti
displaced along $\langle 111 \rangle$ direction), the orthorhombic
structure (space group $Amm2$ with Ti displaced along $\langle
011\rangle$ direction), the tetragonal structure (space group $P4mm$
with Ti displaced along $\langle 001\rangle$ direction) and the cubic
structure (space group $Pm\bar{3}m$ with Ti at the center of oxygen
octahedron). Fig.~\ref{fig:structure}\textbf{a} shows that as electron
doping concentration $n$ increases from 0 to 0.15$e$/f.u., BaTiO$_3$
transitions from the rhombohedral structure to the tetragonal
structure, and finally to the cubic structure.  The critical
concentration is such that the crystal structure of doped BaTiO$_3$
continuously changes from tetragonal to cubic. While the structural
transition from tetragonal to cubic is continuous, the transition from
rhombohedral to orthorhombic is first-order and thus does not show
phonon softening. Furthermore the low electron concentration in the
rhombohedral structure invalidates Migdal's theorem and
electron-phonon coupling can not be calculated within Migdal's
approximation {(Supplementary Note V)}.

Fig.~\ref{fig:structure}\textbf{b} shows $c/a$ ratio and
Ti-O cation displacements $\delta$ as a function of the concentration
$n$ in the range of 0.06-0.14$e$/f.u.. It is evident that the critical
concentration $n_c$ of doped BaTiO$_3$ is 0.10$e$/f.u.  (about
$1.6\times10^{21}$ cm$^{-3}$), at which the polar displacement $\delta$ is just
completely suppressed and the $c/a$ ratio is reduced to unity.
This result is consistent with the previous {  theoretical} studies
~\cite{PhysRevLett.109.247601, Chengliang2019}. {   Experimentally,
  in metallic oxygen-deficient BaTiO$_{3-\delta}$, the low-symmetry polar
  structure can be retained up to an electron concentration of $1.9\times 10^{21}$cm$^{-3}$ (close to the theoretical result)~\cite{PhysRevLett.104.147602,PhysRevB.82.214109}. However, weak localization and/or
  phase separation may exist in oxygen-deficient BaTiO$_3$, depending on sample
quality~\cite{PhysRevLett.104.147602,PhysRevB.84.064125}.}

\subsection*{Electronic structure and phonon properties}

\begin{figure}[t!]
\centering \includegraphics[width=0.9\textwidth]{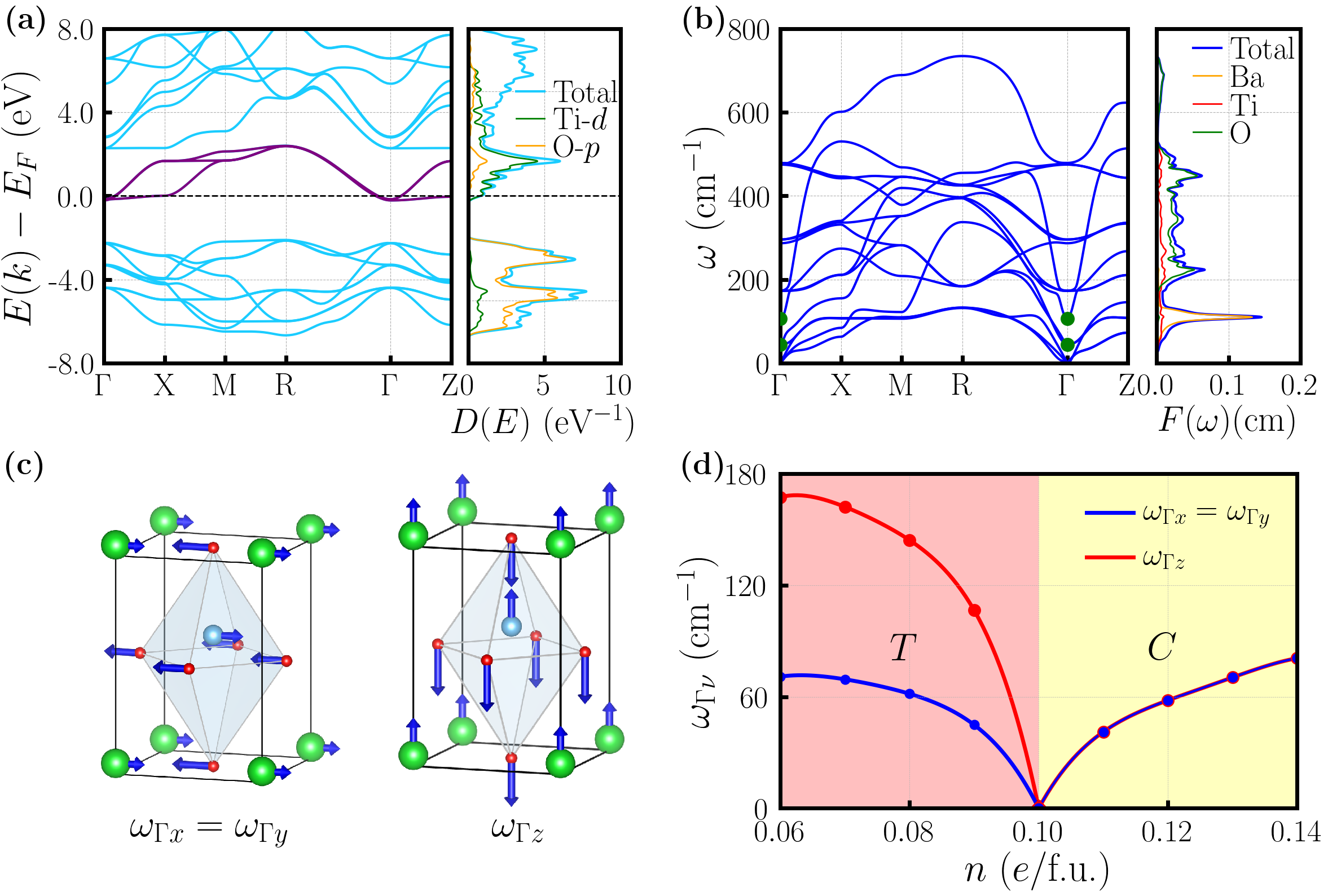}
\caption{\label{fig:e-p}\textbf{ Electronic structure and phonon properties.} \textbf{a}) Electronic band structure and
  density of states of doped BaTiO$_3$ in the tetragonal structure at
  $0.09e$/f.u. concentration. In the electronic band structure,
  the three purple bands are generated by three
  maximally localized Wannier functions that exactly reproduce the
  original Ti $t_{2g}$ bands. In the electronic
  density of states, the blue, green
  and orange curves correspond to total, Ti-$d$ projected and O-$p$
  projected partial densities of states, respectively.
  \textbf{b}) Phonon band structure and
  phonon density of states of doped BaTiO$_3$ in the tetragonal structure
  at $0.09e$/f.u. concentration. In the phonon band structure, 
  the green dots highlight the zone-center
  polar optical phonons. In the phonon density of states, the blue, orange,
  red and green curves correspond to total, Ba-projected, Ti-projected
  and O-projected partial densities of states, respectively.
  \textbf{c}) Vibration modes of the
  zone-center polar optical phonons of doped BaTiO$_3$ in the tetragonal
  structure at $0.09e$/f.u. concentration. The left panel shows that
  the atoms of BaTiO$_3$ are vibrating along the short $a$ axis
  (either $x$-axis or $y$-axis, degenerate due to the tetragonal symmetry).
  The right panel shows that the atoms of BaTiO$_3$ are vibrating along
  the long $c$ axis ($z$-axis). \textbf{d}) The frequencies of the three
  zone-center polar phonons of doped BaTiO$_3$ as a function of electron
  concentration $n$. The critical concentration is at $0.1e$/f.u. where the
  polar phonon frequencies are reduced to zero.}
\end{figure}

Fig.~\ref{fig:e-p}\textbf{a} shows the electronic structure of doped
BaTiO$_3$ in the tetragonal structure at a representative
concentration ($n = 0.09e$/f.u., close to the critical value). Undoped
BaTiO$_3$ is a wide gap insulator.  Electron doping moves the Fermi
level slightly above the conduction band edge of the three Ti $t_{2g}$
orbitals and thus a Fermi surface is formed. We use three Wannier
functions to reproduce the Ti $t_{2g}$ bands, upon which
electron-phonon coupling is calculated. Fig.~\ref{fig:e-p}\textbf{b}
shows the phonon spectrum of doped BaTiO$_3$ in the tetragonal
structure at $0.09e$/f.u. concentration.  We are particularly
interested in the zone-center ($\Gamma$-point) polar optical phonons,
which are highlighted by the green dots in
Fig.~\ref{fig:e-p}\textbf{b}.  The vibrational modes of those polar
phonons are explicitly shown in Fig.~\ref{fig:e-p}\textbf{c}. In the
tetragonal structure of BaTiO$_3$, the two polar phonons with the ion
displacements along $x$ and $y$ directions ($\omega_x$ and $\omega_y$)
are degenerate, while the third polar phonon with the ion
displacements along $z$ direction ($\omega_z$) has higher
frequency. Fig.~\ref{fig:e-p}\textbf{d} shows that electron doping
softens the zone-center polar phonons of BaTiO$_3$ in the tetragonal
structure until it reaches the critical concentration where the three
polar phonon frequencies become zero. With further electron doping,
the polar phonon frequencies of BaTiO$_3$ increase in the cubic
structure.

\subsection*{Electron-phonon coupling and phonon-mediated
  superconductivity}

The continuous polar-to-centrosymmetric phase transition in doped
BaTiO$_3$ is similar to the one in ``ferroelectric-like metals''
proposed by Anderson and Blount~\cite{PhysRevLett.14.217}. They first
argued, later recast by Puggioni and
Rondinelli~\cite{Puggioni2014,Laurita2019}, that inversion symmetry
breaking by collective polar displacements in a metal relies on a weak
coupling between itinerant electrons and soft phonons responsible for
removing inversion symmetry. According to this argument, one would
expect that across the polar-to-centrosymmetric phase transition, the
soft polar phonons are not strongly coupled to itinerant electrons in
doped BaTiO$_3$. {  In order to quantify the strength of electron-phonon
coupling and make quantitative comparison, we introduce the
mode-resolved electron-phonon coupling $\lambda_{\textbf{q}\nu}$ and
around-zone-center branch-resolved electron-phonon coupling
$\lambda_{\nu}$:
\begin{equation}
\label{eqn:lambda_qv} \lambda_{\textbf{q}\nu} = \frac{1}{\pi
  N_F}\frac{\textrm{Im}\Pi_{\textbf{q}\nu}}{\omega^2_{\textbf{q}\nu}} \textrm{ and }
\lambda_{\nu} = \frac{\int_{|\mathbf{q}| < q_c} d\mathbf{q}\lambda_{\textbf{q}\nu}}{\int_{|\mathbf{q}|<q_c} d\mathbf{q}}
\end{equation}
where $\textrm{Im}\Pi_{\textbf{q}\nu}$ is the imaginary part of
electron-phonon self-energy, $\omega_{\textbf{q}\nu}$ is the phonon
frequency, $N_F$ is the density of states at the Fermi level and
$q_c$ is a small phonon momentum. The reason we define $\lambda_{\nu}$
within $|\textbf{q}|< q_c$ is because: 1) exactly at the zone-center
$\Gamma$ point, the
acoustic phonon frequency is zero and thus the contribution from the acoustic
mode is ill-defined at $\Gamma$ point; 2) when $q_c$ is sufficiently small,
there are no phonon band crossings within $|\textbf{q}| < q_c$ and
hence each branch $\nu$ can be assigned to a well-defined phonon mode
(for a general $\textbf{q}$ point, it is not
trivial to distinguish which phonon band corresponds to polar modes and
which to other optical modes). We choose $q_c = 0.05 \frac{\pi}{a}$ where
$a$ is the lattice constant (the qualitative conclusions do not depend on the
choice of $q_c$, as long as no phonon band crossings occur within
$|\textbf{q}| < q_c$).}

Fig.~\ref{fig:superconducting}\textbf{a} and \textbf{b} show the
imaginary part of electron-phonon self-energy
$\textrm{Im}\Pi_{\textbf{q}\nu}$ for each phonon mode $\textbf{q}\nu$
of doped BaTiO$_3$ along a high-symmetry path (panel \textbf{a}
corresponds to $0.09e$/f.u. doping in a tetragonal structure
and panel \textbf{b} corresponds to $0.11e$/f.u. doping in a
cubic structure). The point size is proportional to
$\textrm{Im}\Pi_{\textbf{q}\nu}$~\cite{double}. Our calculations find
that, contrary to Anderson/Blount's weak coupled electron
mechanism~\cite{PhysRevLett.14.217}, the phonon bands associated with
the zone-center polar phonons have the strongest coupling to itinerant
electrons, while the couplings of other phonon bands are weaker.
{  Specifically, in the case of 0.09$e$/f.u. doping:
\begin{eqnarray}
  \label{eqn:lambdapolar1} \lambda_{\textrm{acoustic}} = \sum_{\nu=1-3}\lambda_{\nu} = 3.83 \\\nonumber
  \lambda_{\textrm{polar}} = \sum_{\nu=4-6}\lambda_{\nu} = 10.92 \\\nonumber
  \lambda_{\textrm{others}} = \sum_{\nu=7-15}\lambda_{\nu} = 0.53
\end{eqnarray}  
and in the case of 0.11$e$/f.u. doping:
\begin{eqnarray}
  \label{eqn:lambdapolar2} \lambda_{\textrm{acoustic}} = \sum_{\nu=1-3}\lambda_{\nu} = 0.27 \\\nonumber
  \lambda_{\textrm{polar}} = \sum_{\nu=4-6}\lambda_{\nu} = 5.58 \\\nonumber
  \lambda_{\textrm{others}} = \sum_{\nu=7-15}\lambda_{\nu} = 0.11
\end{eqnarray}  
In both cases, $\lambda_{\textrm{polar}}$ is larger than
$\lambda_{\textrm{acoustic}}$ and $\lambda_{\textrm{others}}$.} An
intuitive picture for the strong coupling is that in doped BaTiO$_3$,
the soft polar phonons involve the cation displacements of Ti and O
atoms, and in the meantime itinerant electrons derive from Ti-$d$
states which hybridize with O-$p$ states {  (see
  Supplementary Note XIV for an alternative demonstration of this
  strong coupling)}.  This is in contrast to the textbook example of
  polar metals LiOsO$_3$ where the soft polar phonons involve Li
displacements while the metallicity derives from Os and O orbitals
~\cite{Laurita2019}. {  More quantitatively, we find
  $\lambda_{\textrm{polar}} = 0.50$ 
  for LiOsO$_3$, which is substantially
  smaller than $\lambda_{\textrm{polar}}$ of about $5\sim 10$ for
  doped BaTiO$_3$.
  In short, because the itinerant
  electrons and polar phonons are associated with the same atoms in
  doped BaTiO$_3$, the coupling is strong, while in LiOsO$_3$ the
  itinerant electrons and polar phonons involve different atoms and
  thus the coupling is weak.} As a consequence of the strong
interaction between the polar phonons and itinerant electrons, we
expect that the total electron-phonon coupling of doped BaTiO$_3$ can
be increased by softening the polar phonons across the structural
phase transition.

Fig.~\ref{fig:superconducting}\textbf{c}
shows the total electron-phonon spectral function $\alpha^2F(\omega)$ and
accumulative electron-phonon coupling $\lambda(\omega)$
of doped BaTiO$_3$ at $0.09e$/f.u. and $0.11e$/f.u. concentrations.
$\alpha^2F(\omega)$ is defined as:
\begin{equation}
  \label{eqa2f} \alpha^2F(\omega) = \frac{1}{2} \sum_{\nu}
  \int \frac{d\textbf{q}}{\Omega_{\textrm{BZ}}} \omega_{\textbf{q}\nu}\lambda_{\textbf{q}\nu}\delta(\omega-\omega_{\textbf{q}\nu}) 
\end{equation}
where $\Omega_{\textrm{BZ}}$ is the volume of phonon Brillouin
zone. With
$\alpha^2F(\omega)$, it is easy to calculate the accumulative
electron-phonon coupling $\lambda(\omega)$:
\begin{equation}
  \label{eqlambda} \lambda(\omega) = 2\int^{\omega}_{0} \frac{\alpha^2F(\nu)}{\nu}d\nu
\end{equation}
The total electron-phonon coupling $\lambda$ is obtained by taking the
upper bound $\omega$ to $\infty$ in Eq.~(\ref{eqlambda}).
The green shades are $\alpha^2F(\omega)$ and the dashed lines are
the corresponding cumulative electron-phonon coupling. The total
electron-phonon coupling $\lambda$ of doped BaTiO$_3$ in the
tetragonal structure at $0.09e$/f.u. concentration is 0.61, while that
in the cubic structure at $0.11e$/f.u. concentration is 0.50. Both $\lambda$ are
sufficiently large to induce phonon-mediated superconductivity with
measurable transition temperature.
Fig.~\ref{fig:superconducting}\textbf{d} shows the total
electron-phonon coupling $\lambda$ of doped BaTiO$_3$ for a range of
electron concentrations (exactly at the critical concentration, we
find some numerical instabilities and divergence in the
electron-phonon calculations, rendering the result unreliable).
An increase of $\lambda$ around the critical
concentration is evident, consistent with the strong coupling between
the soft polar phonons and itinerant electrons in doped BaTiO$_3$.

Based on the electron-phonon spectrum $\alpha^2F(\omega)$, we use a
three-orbital Eliashberg equation ( Supplementary Note III) to
calculate the superconducting gap $\Delta(T)$ and estimate the
superconducting transition temperature $T_c$ as a function of electron
concentration. Because the three Ti $t_{2g}$ orbitals become identical
at the critical concentration, when solving the three-orbital
Eliashberg equation, we set Morel-Anderson pseudopotential
$\mu^{*}_{ij}$ to be 0.1 for each orbital pair
(i.e. $i,j=1,2,3$)~\cite{PhysRev.125.1263}.
Fig.~\ref{fig:superconducting}\textbf{e} shows the superconducting gap
$\Delta(T)$ of doped BaTiO$_3$ as a function of temperature $T$ at two
representative concentrations (0.09$e$/f.u. in the tetragonal
structure and 0.11$e$/f.u. in the cubic structure). Since both
concentrations are close to the critical value, the three Ti $t_{2g}$
orbitals are almost degenerate in doped BaTiO$_3$. For clarification,
we show the superconducting gap of one orbital for each
concentration. From the Eliashberg equation, we find that at
0.09$e$/f.u. concentration, $\Delta(T=0)=0.27$ meV and $T_c=1.75$ K;
and at 0.11$e$/f.u. concentration, $\Delta(T=0)=0.11$ meV and
$T_c=0.76$ K.  Thus $\Delta (T=0)/(k_BT_c) = 1.79$ at
0.09$e$/f.u. concentration and 1.68 at 0.11$e$/f.u. concentration,
both close to the BCS prediction of 1.77.
Fig.~\ref{fig:superconducting}\textbf{f} shows the estimated
superconducting transition temperature $T_c$ of doped BaTiO$_3$ for a
range of electron concentrations.  $T_c$ notably exhibits a dome-like
feature as a function of electron concentration.  The origin of the
superconducting ``dome'' is that the electron-phonon coupling of doped
BaTiO$_3$ is increased by the softened polar phonons around the
critical concentration. When the electron concentration is away from
the critical value, the polar phonons are ``hardened'' (i.e. phonon
frequency increases) and the electron-phonon coupling of doped
BaTiO$_3$ decreases.  We note that the estimated $T_c$ strongly
depends on $\mu^{*}_{ij}$. Therefore in the inset of
Fig.~\ref{fig:superconducting}\textbf{f}, we study doped BaTiO$_3$ at
a representative concentration (0.09$e$/f.u.) and calculate its
superconducting transition temperature $T_c$ as a function of
$\mu^{*}_{ij}$.  As $\mu^{*}_{ij}$ changes from 0 to 0.3, the
estimated $T_c$ decreases from 9.3 K to 0.4 K, the lowest of which
(0.4 K) is still measurable in experiment~\cite{PhysRevX.3.021002}.
{  We make two comments here: 1) The superconducting
  transition temperature is only an estimation due to the uncertainty
  of Morel-Anderson pseudopotential $\mu^{*}_{ij}$ and other technical
  details. But the picture of an increased electron-phonon coupling
  around the structural phase transition in doped BaTiO$_3$ is
  robust. 2) Experimentally in Sr$_{1-x}$Ca$_x$TiO$_3$, the optimal
  doping for superconductivity is larger than the ``ferroelectric''
  critical concentration~\cite{Rischau2017}, while in our calculations
  of doped BaTiO$_3$, the two critical concentrations (one for optimal
  superconducting $T_c$ and the other for suppressing polar
  displacements) just coincide due to polar phonon softening and an
  increased electron-phonon coupling. Comparison of these two
  materials implies that the microscopic mechanism for
  superconductivity in doped SrTiO$_3$ is probably not purely
  phonon-mediated. }

\subsection*{Crystal symmetry and acoustic phonons}

\begin{figure}[t]
\centering
\includegraphics[width = 0.9\textwidth]{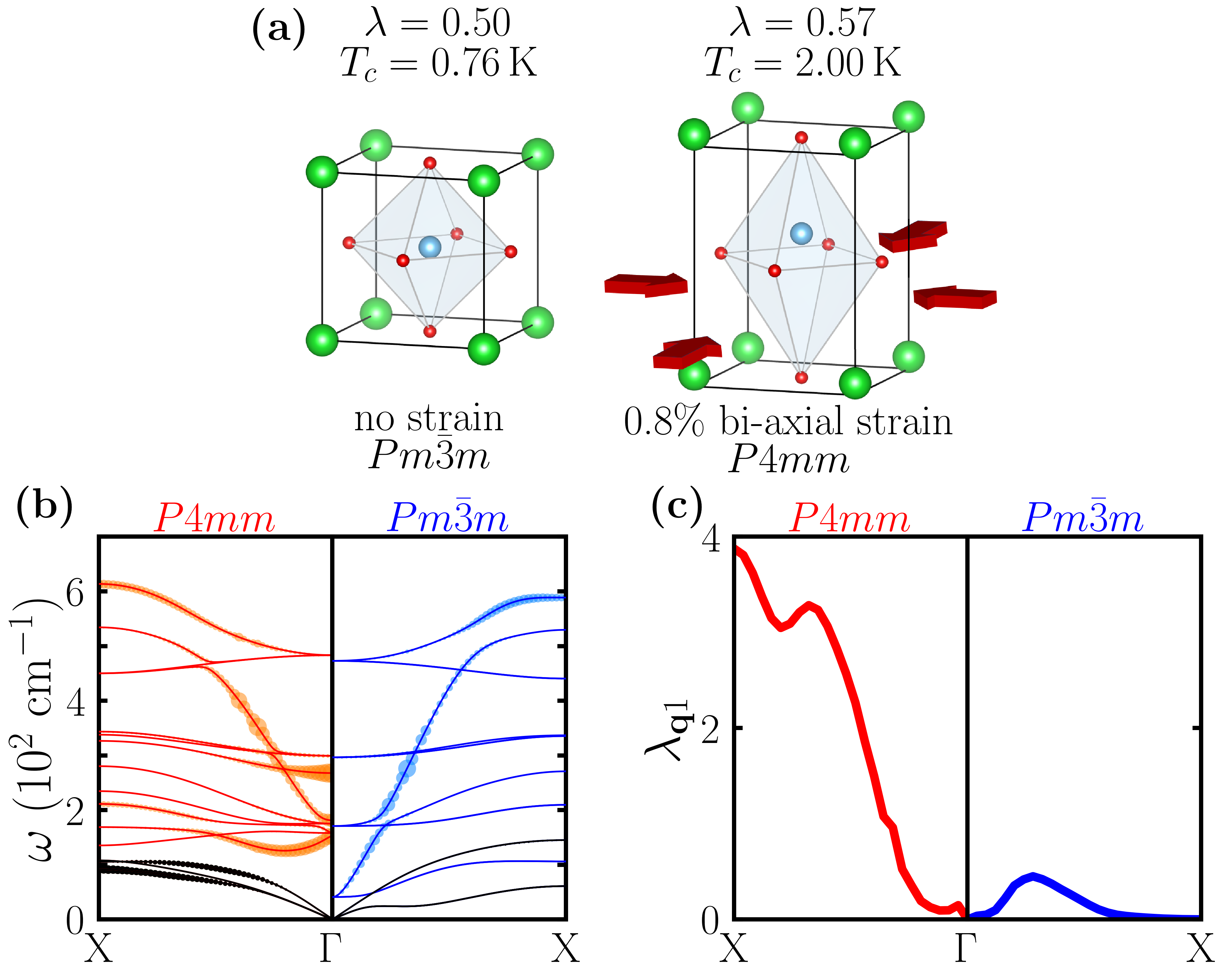}
\caption{\label{fig:difference}\textbf{ Crystal symmetry and acoustic phonons.} \textbf{a}) Doped BaTiO$_3$ at
  $0.11e$/f.u. concentration. Left is the cubic crystal structure of
  BaTiO$_3$ with no strain (space group $Pm\bar{3}m$) and right is the
  polar tetragonal crystal structure of BaTiO$_3$ under 0.8\% bi-axial
  strain (space group $P4mm$). \textbf{b}) The imaginary part of the
  electron-phonon self-energy $\textrm{Im}\Pi_{\textbf{q}\nu}$ for
  each phonon mode of doped BaTiO$_3$, at 0.11$e$/f.u. in the $P4mm$
  structure (red, left) and in the $Pm\bar{3}m$ structure (blue,
  right).  The point size is proportional to
  $\textrm{Im}\Pi_{\textbf{q}\nu}$.  The largest point corresponds to
  $\textrm{Im}\Pi_{\textbf{q}\nu}$ = 3.8 meV.  \textbf{c})
  {   Mode-resolved electron-phonon coupling
    $\lambda_{\textbf{q}\nu}$ for each phonon mode of doped BaTiO$_3$,
    at 0.11$e$/f.u. in the $P4mm$ structure (red, left) and in the
    $Pm\bar{3}m$ structure (blue, right).  The point size is proportional to
    $\lambda_{\textbf{q}\nu}$. The largest point corresponds to
    $\lambda_{\textbf{q}\nu}$ = 5.1.}}
\end{figure}

We note that in Fig.~\ref{fig:superconducting}\textbf{a} and
\textbf{b}, in addition to the large $\textrm{Im}\Pi_{\textbf{q}\nu}$
in the polar optical phonon bands, there is also sizable
$\textrm{Im}\Pi_{\textbf{q}\nu}$ in the acoustic phonon bands (from
$\Gamma$ to X) in the tetragonal structure at $0.09e$/f.u.
concentration. {  Since the mode-resolved electron-phonon
  coupling $\lambda_{\textbf{q}\nu} \propto
  \textrm{Im}\Pi_{\textbf{q}\nu}/\omega^2_{\textbf{q}\nu}$, the small
  frequency of acoustic phonons can lead to a substantial
  $\lambda_{\textbf{q}\nu}$, given a sizable
  $\textrm{Im}\Pi_{\textbf{q}\nu}$.}  However, in the cubic structure
at $0.11e$/f.u. concentration, $\textrm{Im}\Pi_{\textbf{q}\nu}$ in the
acoustic phonon bands almost vanishes from $\Gamma$ to X. To exclude
that the concentration difference may have an effect, we perform a
 {numerical experiment}: we start from the cubic structure doped
at $0.11e$/f.u. concentration (space group $Pm\bar{3}m$), and then we
impose a slight {  (001)} compressive bi-axial 0.8\% strain
{   by fixing the two in-plane lattice constants ($a$ and
  $b$) to a smaller value}.  This compressive strain makes the crystal
structure of doped BaTiO$_3$ tetragonal and polar (space group
$P4mm$). Fig.~\ref{fig:difference}\textbf{a} shows the optimized
crystal structures of the two doped BaTiO$_3$. For doped BaTiO$_3$ at
0.11$e$/f.u. concentration, without strain, the ground state structure
is cubic and the optimized lattice constant $a$ is 3.972 \AA;
under a 0.8\% biaxial {   (001)}
compressive strain, the ground state structure
becomes tetragonal with the in-plane lattice constants $a$ and $b$ being
fixed at 3.940 \AA~and the optimized long lattice constant $c$ being 4.019
\AA. We find that the total electron-phonon coupling $\lambda$
increases from 0.50 in the $Pm\bar{3}m$ structure to 0.57 in the
$P4mm$ structure.
Fig.~\ref{fig:difference}\textbf{b} and
Fig.~\ref{fig:difference}\textbf{c} compare the imaginary part of the
electron-phonon self-energy $\textrm{Im}\Pi_{\textbf{q}\nu}$ and the
mode-resolved electron-phonon coupling $\lambda_{\textbf{q}\nu}$ along
the $\Gamma\to X$ path for the two doped BaTiO$_3$. Similar to
Fig.~\ref{fig:superconducting}\textbf{a} and \textbf{b}, we find that
there is a notable difference in $\textrm{Im}\Pi_{\textbf{q}\nu}$ from
the acoustic phonon bands. The difference in
$\textrm{Im}\Pi_{\textbf{q}\nu}$ is further ``amplified'' by the low
phonon frequencies $\omega_{\textbf{q}\nu}$, {  which
  results in
\begin{eqnarray}
\label{eqn:compareacoustic} \textrm{without epitaxial strain}\phantom{10} \lambda_{\textrm{acoustic}} = 0.27\\\nonumber
  \textrm{under 0.8\% (001) compressive strain} \phantom{10} \lambda_{\textrm{acoustic}} = 4.45
\end{eqnarray}
At the same time, we find that for polar modes,
\begin{eqnarray}
\label{eqn:comparepolar} \textrm{without epitaxial strain}\phantom{10} \lambda_{\textrm{polar}} = 5.58\\\nonumber
  \textrm{under 0.8\% (001) compressive strain} \phantom{10} \lambda_{\textrm{polar}} = 3.21
\end{eqnarray}
This shows that under 0.8\% (001) compressive strain, 
$\lambda_{\textrm{polar}}$ remains substantial
(albeit reduced by about 40\%), 
but $\lambda_{\textrm{acoustic}}$ is increased by one order of magnitude, which
altogether leads to an enhancement of the total electron-phonon
coupling $\lambda$.}  Note that in the  {numerical experiment},
the two doped BaTiO$_3$ have exactly the same electron concentration,
indicating that the additional increase in
$\textrm{Im}\Pi_{\textbf{q}\nu}$ of the acoustic phonons
arises solely from the crystal
structure difference.  {   A possible explanation, which is
  based on our calculations, is that in the cubic structure, some
  electron-phonon vertices $g_{ij}^{\nu}(\textbf{k},\textbf{q})$ are
  exactly equal to zero because some atoms are frozen in the acoustic
  phonons, while in the low-symmetry structure, those
  $g_{ij}^{\nu}(\textbf{k},\textbf{q})$ become non-zero. Because
  $\textrm{Im}\Pi_{\textbf{q}\nu}\propto|g^{\nu}_{ij}(\textbf{k},\textbf{q})|^2$
  ~\cite{PhysRevB.87.024505, Giustino2017}, this leads to an increase
  in $\textrm{Im}\Pi_{\textbf{q}\nu}$. In addition, the frequencies of
  acoustic phonon modes $\omega_{\textbf{q}\nu}$ are very small and
  $\lambda_{\textbf{q}\nu}\propto\textrm{Im}\Pi_{\textbf{q}\nu}/\omega^2_{\textbf{q}\nu}$,
  therefore even a slight increase in $\textrm{Im}\Pi_{\textbf{q}\nu}$
  results in a substantial enhancement in $\lambda_{\textbf{q}\nu}$
  (see Supplementary Note XVI for the demonstration of a specific
  acoustic phonon).}  Our  {numerical experiment} also
{  implies} that in doped BaTiO$_3$, when the electron
concentration is close to the critical value, a small
{  (001)} compressive strain that lowers the crystal
symmetry may also enhance its superconducting transition temperature
{  due to the increased electron-phonon coupling}, similar
to doped SrTiO$_3$~\cite{Ahadieaaw0120,Russell2019}.

\section*{Discussion}

Finally we discuss possible experimental verification. Chemical doping
~\cite{10.4236/msa.2011.27099,IANCULESCU201110040,MORRISON20011205,Takahashi2017,Zhang_2019,PhysRevLett.104.147602,PhysRevB.82.214109}
and epitaxial
strain~\cite{schlom_chen_fennie_gopalan_muller_pan_ramesh_uecker_2014,MARTIN2012199}
have been applied to ferroelectric materials such as
BaTiO$_3$. La-doped BaTiO$_3$ has been experimentally synthesized.
High-temperature transport measurements show that
Ba$_{1-x}$La$_x$TiO$_3$ exhibits polar metallic behaviour but {  
ultra-low-temperature transport measurements are yet to be 
performed
~\cite{IANCULESCU201110040,10.4236/msa.2011.27099,Takahashi2017,Zhang_2019,MORRISON20011205}. 
  We note that La doping in BaTiO$_3$ may result in some chemical
  disorder. While the randomness of La distribution in
  La$_x$Ba$_{1-x}$TiO$_3$ may affect the transport properties in the
  normal state, Anderson's theorem asserts that superconductivity in a
  conventional superconductor is robust with respect to non-magnetic
  disorder in the host material~\cite{ANDERSON195926}. As a
  consequence, the superconducting transition temperature $T_c$ of a
  conventional superconductor barely depends on the randomness of
  defects. In our case, the superconductivity in doped BaTiO$_3$ is
  phonon-mediated (i.e. conventional) and La is a non-magnetic dopant.
  Therefore Anderson's theorem applies and we expect that even if
  chemical disorder may arise in actual experiments, it does not
  affect the superconducting properties of doped BaTiO$_3$. In
  addition, we perform supercell calculations which include real La
  dopants (Supplementary Note VIII).  We find that even in the
  presence of real La atoms, the conduction electrons on Ti atoms are
  almost uniformly distributed in La$_x$Ba$_{1-x}$TiO$_3$.}  Since our
simulation does not consider dopants explicitly, a more desirable
doping method is to use electrostatic carrier doping
~\cite{doi:10.1063/1.3669402,Eyvazov2013,Ye1193}, which does not
involve chemical dopants and has been successfully used to induce
superconductivity in KTaO$_3$~\cite{Ueno2011}. {   We
  clarify two points concerning the electrostatic doping method. 1)
  The electrostatic gating by ionic liquid can achieve a
  two-dimensional carrier density as high as $8\times10^{14}$
  cm$^{-2}$~\cite{https://doi.org/10.1002/adfm.200801633}.  The
  induced electrons are usually confined in a narrow region that is a
  few nanometers from the surface/interface, which leads to an
  effective three-dimensional carrier density of about
  $1\times 10^{21}\sim 5\times 10^{21}$cm$^{-3}$~\cite{Ueno2011,PhysRevLett.102.216804}. In our current study, the critical
  concentration of doped BaTiO$_3$ is about $1.6\times
  10^{21}$cm$^{-3}$, which is feasible by this approach.  2) While the
  electrostatic doping method induces the carriers in the
  surface/interface area, we show that our results on bulk doped
  BaTiO$_3$ can still be used as a guidance to search for superconductivity in
  the surface area of BaTiO$_3$.  We perform calculations of
  Pt/BaTiO$_3$ interface (Supplementary Note X) and find that
  just in the second unit cell of BaTiO$_3$ from the interface, the
  Ti-O displacement saturates and a bulk-like region emerges with
  almost uniform cation displacements. In addition, we calculate the
  electron-phonon properties of bulk KTaO$_3$ at 0.14$e$/f.u. doping
  (based on the experiment~\cite{Ueno2011}) (Supplementary Note
  XV).  We find that the total electron-phonon coupling of KTaO$_3$
  at 0.14$e$/f.u. doping is 0.36.  Using McMillian equation (take
  $\mu^{*}=0.1$) as a rough estimation of superconducting transition
  temperature $T_c$, we obtain a $T_c$ of about 68 mK, which is in
  reasonable agreement with the experimental value of 50 mK. While
  there is definitely room for improvement, our results demonstrate that
  for a given target material, its desirable
  bulk electron-phonon property can point to the right direction in
  which superconductivity is found in surface/interface regions.}

In summary, we use first-principles calculations to demonstrate a
large modulation of electron-phonon coupling and an emergent
superconducting ``dome'' in $n$-doped BaTiO$_3$.
Contrary to Anderson/Blount's weak electron coupling mechanism for
``ferroelectric-like metals''~\cite{PhysRevLett.14.217,Laurita2019,Puggioni2014}, our calculations find
that the soft polar phonons are strongly coupled to itinerant electrons
across the polar-to-centrosymmetric phase transition
in doped BaTiO$_3$ and as a consequence, the total electron-phonon coupling
increases around the critical concentration.
In addition, we find that lowering the crystal symmetry of doped
BaTiO$_3$ by imposing epitaxial strain can also
increase the electron-phonon coupling via a sizable coupling between
acoustic phonons and itinerant electrons.
Our work provides an experimentally viable method to modulating
electron-phonon coupling and inducing phonon-mediated
superconductivity in doped strong ferroelectrics. Our
results indicate that the weak electron coupling mechanism for
``ferroelectric-like metals''~\cite{PhysRevLett.14.217,Laurita2019,Puggioni2014} is not necessarily present in
doped strong ferroelectrics. We hope that our predictions
will stimulate experiments on doped ferroelectrics and {   search
for the phonon-mediated superconductivity that is predicted in our calculations.}

\section*{Methods}

We perform first-principles calculations by using density functional
theory (DFT)~\cite{PhysRev.136.B864,PhysRev.140.A1133,
  RevModPhys.73.515, RevModPhys.84.1419} as implemented in the Quantum
ESPRESSO package~\cite{Giannozzi_2009}. We use norm-conserving
pseudo-potentials~\cite{VANSETTEN201839} with local density approximation as the
exchange-correlation functional. For electronic structure
calculations, we use an energy cutoff of 100 Ry. {  We optimize
both cell parameters and internal coordinates in atomic relaxation.
In the strain calculations, the in-plane lattice constants are fixed
while the out-of-plane lattice constant and internal coordinates
are fully optimized.}
The electron
Brillouin zone integration is performed with a Gaussian smearing of
0.005 Ry over a $\Gamma$-centered \textbf{k} mesh of $12\times 12
\times 12$. The threshold of total energy convergence is $10^{-7}$~Ry;
self-consistency convergence is $10^{-12}$ Ry; force convergence is
$10^{-6}$ Ry/Bohr and pressure convergence for variable cell is 0.5
kbar. For phonon calculations, we use density functional perturbation
theory (DFPT)~\cite{RevModPhys.73.515} as implemented in the Quantum
ESPRESSO package~\cite{Giannozzi_2009}. The phonon Brillouin zone
integration is performed over a \textbf{q} mesh of $6 \times 6
\times6$.  For the calculations of electron-phonon coupling and
superconducting gap, we use maximally localized Wannier functions and
Migdal-Eliashberg theory, as implemented in the
Wannier90~\cite{Pizzi2020} and EPW code~\cite{epw}. The Fermi surface
of electron-doped BaTiO$_3$ is composed of three Ti $t_{2g}$
orbitals. We use three maximally localized Wannier functions to
reproduce the Fermi surface. The electron-phonon matrix elements
$g_{ij}^{\nu}(\textbf{k},\textbf{q})$ are first calculated on a coarse
$12\times 12\times 12$ \textbf{k}-grid in the electron Brillouin zone
and a coarse $6\times 6 \times 6$ \textbf{q}-grid in the phonon
Brillouin zone, and then are interpolated onto fine grids via
maximally localized Wannier functions.  The fine electron and phonon
grids are both $50\times50\times 50$. We check the convergence on the
electron \textbf{k}-mesh, phonon \textbf{q}-mesh and Wannier
interpolation and no significant difference is found by using a denser
mesh.  Details can be found in Supplementary Note IV).  We
solve a three-orbital Eliashberg equation to estimate the
superconducting transition temperature $T_c$ (   Supplementary
Note III).

We only use Eliashberg equation when electron doping concentration is
high enough so that $\lambda T_D/T_F < 0.1$ and Migdal's theorem is
valid~\cite{Migdal} ($\lambda$ is electron-phonon coupling, $T_D$ is Debye
temperature and $T_F$ is Fermi temperature). Validation test of
Migdal's theorem is shown in Supplementary Note V.

We solve a three-orbital Eliashberg equation to estimate the
superconducting transition temperature $T_c$. This method is compared
to McMillan Equation. Details of Eliashberg Equation and McMillan
Equation can be found in Supplementary Note III.

\clearpage
\newpage

\section*{Data availability}
The data that support the findings of this study are available from
the corresponding author upon reasonable request.

\section*{Code availability}

The electronic structure calculations were performed using
the open-source code Quantum Espresso~\cite{Giannozzi_2009}.
Quantum Espresso package is freely
distributed on academic use under the Massachusetts Institute of Technology
(MIT) License.

\clearpage
\newpage

\begin{figure}[t]
\centering
\includegraphics[width=0.9\textwidth]{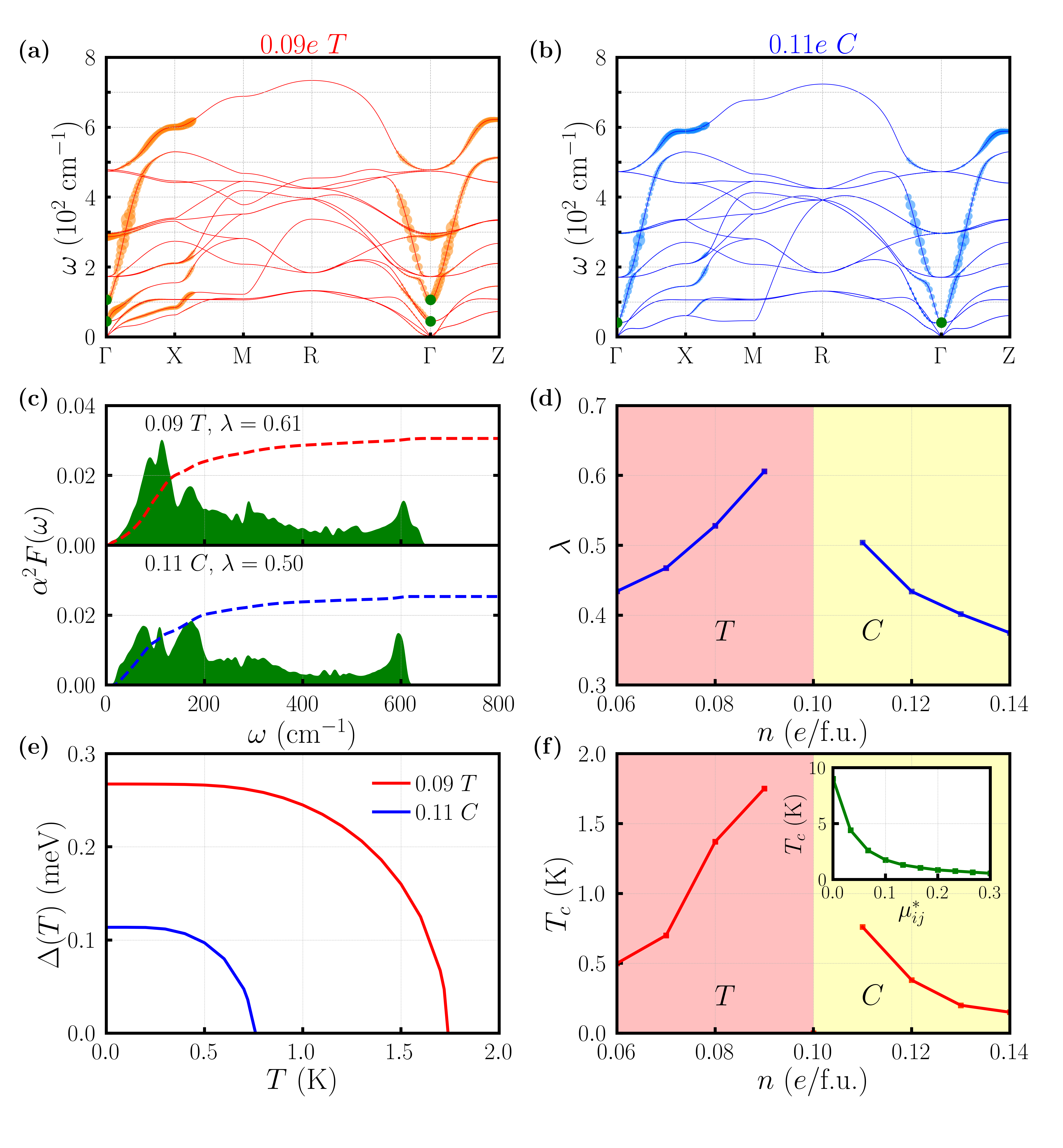}
\end{figure}

\begin{figure}
    \centering
\caption{\label{fig:superconducting}\textbf{ Electron-phonon coupling and phonon-mediated superconductivity.} \textbf{a}) The imaginary part of
  the electron-phonon self-energy $\textrm{Im}\Pi_{\textbf{q}\nu}$ for
  each phonon mode of doped BaTiO$_3$ at $0.09e$/f.u. concentration
  (tetragonal structure $T$). The point size is proportional to
  $\textrm{Im}\Pi_{\textbf{q}\nu}$. The largest point corresponds to
  $\textrm{Im}\Pi_{\textbf{q}\nu}$ = 4.6 meV. The green dots highlight
  the zone-center polar optical phonons.  \textbf{b}) The imaginary
  part of the electron-phonon self-energy
  $\textrm{Im}\Pi_{\textbf{q}\nu}$ for each phonon mode of doped
  BaTiO$_3$ at $0.11e$/f.u. concentration (cubic structure $C$).  The
  point size is proportional to $\textrm{Im}\Pi_{\textbf{q}\nu}$.  The
  largest point corresponds to $\textrm{Im}\Pi_{\textbf{q}\nu}$ = 3.2
  meV. The green dots highlight the zone-center polar optical phonons.
  \textbf{c}) Electron-phonon spectral function $\alpha^2F(\omega)$
  and accumulative electron-phonon coupling $\lambda(\omega)$ of doped
  BaTiO$_3$ at $0.09e$/f.u. and $0.11e$/f.u. concentration. The total
  electron-phonon coupling $\lambda$ is 0.61 for the former and 0.50
  for the latter. \textbf{d}) Total electron-phonon coupling $\lambda$
  of doped BaTiO$_3$ as a function of electron concentration
  $n$. \textbf{e}) Superconducting gap $\Delta$ of doped BaTiO$_3$ as
  a function of temperature $T$ at 0.09 $e$/f.u.  concentration (red)
  and at 0.11 $e$/f.u. concentration (blue), calculated by the
  three-orbital Eliashberg equation. The Morel-Anderson
  pseudopotential $\mu^*_{ij} =0.1$ is used for each orbital pair.
  \textbf{f}) Superconducting transition temperature $T_c$ of doped
  BaTiO$_3$ calculated by the Eliashberg equation as a function of
  electron concentration $n$.  The inset shows $T_c$ of BaTiO$_3$ in
  the tetragonal structure at $0.09e$/f.u.  concentration as a
  function of Morel-Anderson pseudopotential $\mu_{ij}^{*}$.}
\end{figure}

\clearpage
\newpage

\begin{acknowledgments}
We acknowledge useful discussion with Kevin Garrity,
Jia Chen and Jin Zhao.  H.C. is supported by the National Natural
Science Foundation of China under Project No. 11774236 and NYU
University Research Challenge Fund.  J.M. is supported by the Student
Research Program in Physics of NYU Shanghai. NYU high performance
computing at Shanghai, New York and Abu Dhabi campuses provide the
computational resources.
\end{acknowledgments}

\section*{Author Contributions}
J.M. performed the first-principles calculations. R.Y. wrote the code of the Eliashberg Equations solver. H.C. supervised the study. All authors contributed to discussions and writing the manuscript.

\section*{Competing interests}
The authors declare no competing interests.
\end{document}